\documentstyle[aps,pra,multicol,psfig]{revtex}
\begin{document}
\draft

\title{A simple model of DNA denaturation and mutually
avoiding walks statistics}
\author{Marco Baiesi$^{1}$, 
Enrico Carlon$^{2}$, 
Enzo Orlandini$^{1}$ 
and Attilio L. Stella$^{1}$}
\address{$^{1}$INFM - Dipartimento di Fisica, Universit\`a di Padova,
I-35131 Padova, Italy \\
$^{2}$Theoretische Physik, Universit\"at des Saarlandes, 
D-66041 Saarbr\"ucken, Germany
}

\date{\today}
\maketitle

\begin{abstract}
Recently Garel, Monthus and Orland ({\it Europhys. Lett.} {\bf 55}, 
132 (2001)) considered a model of DNA denaturation in which excluded 
volume effects within each strand are neglected, while mutual avoidance 
is included. 
Using an approximate scheme they found a first order denaturation.
We show that a first order transition for this model follows from 
exact results for the statistics of two mutually avoiding random walks, whose 
reunion exponent is $c > 2$, both in two and three dimensions. Analytical 
estimates of $c$ due to the interactions with other denaturated loops, as well
as numerical calculations, indicate that the transition is even
sharper than in models where excluded volume effects are fully incorporated.
The probability distribution of distances between homologous base pairs
decays as a power law at the transition.
\end{abstract}

\pacs{PACS numbers: 87.14.Gg 05.70.Jk 05.70.Fh 87.15.Aa}

\begin{multicols}{2} \narrowtext
\section{Introduction}
\label{sec:intro}

Simple models of DNA thermal denaturation have attracted a lot of attention 
for a long time \cite{Pola66,Fish66,Fish84,Peyr89}. 
Although it was pointed out rather early that excluded volume effects 
may be crucial in determining the order of the denaturation transition 
\cite{Fish66}, it was only very recently that such effects were
taken into account both at analytical and numerical level 
\cite{Caus00,Kafr00,Carl02,Baie02}.

Even once accepted excluded volume effects as a key factor in determining the
nature of the transition, one can still be interested in understanding how the 
relaxation of self-avoidance constraints influences the overall behavior of the 
system. As this can be done in different ways, learning in detail 
about the effects of different geometrical exclusion constraints can help in 
conceiving better models or more efficient approximations.

The main general aim of this paper is that of producing a contribution to
the debate on these general issues. In particular we analyze the model of 
DNA denaturation recently introduced and studied by Garel, Monthus and 
Orland (GMO) \cite{Gare01}. Among other things, we show that this model
is a rather ideal context in which a recently proposed representation
of DNA in terms of block copolymer networks \cite{Baie02} can be applied.

The Hamiltonian for the GMO model is \cite{Gare01}:
\begin{eqnarray}
 H &=&  \frac{1}{2} \sum_{i=1,2} \int_0^N ds
\left( \frac{d {\vec r}_i (s)}{d s} \right)^2
+ \int_0^N ds \ V({\vec r}_{12} (s)) \nonumber \\
&+& g \int_0^N ds \int_0^N ds' \ \delta ({\vec r}_1 (s) - {\vec r}_2 (s'))
\label{hamin}
\end{eqnarray}
where ${\vec r}_1 (s)$, ${\vec r}_2(s)$ describe the conformations of the two 
strands as functions of the curvilinear coordinate $s$ and 
$V({\vec r}_{12}(s))$ is the attractive interaction between homologous 
base pairs (${\vec r}_{12}(s) \equiv {\vec r}_1 (s) - {\vec r}_2 (s)$). 
The last term on the r.h.s. of Eq. (\ref{hamin}) is due to excluded 
volume interactions, which act only between the two chains and not within 
a single chain. Thus the self-avoidance constraint is relaxed and each 
isolated strand follows a simple random walk (RW) statistics. 

The authors of Ref. \cite{Gare01} simplified further the Hamiltonian 
(\ref{hamin}) by approximating the excluded volume term by a long range 
interaction between the strands, which allows for an analytical solution of 
the problem. Within such approximation, the denaturation transition turns 
out to be of first order type \cite{Gare01}, 
as in models where excluded volume effects are fully incorporated 
\cite{Caus00,Kafr00,Carl02,Baie02}.

Our purpose here is to clarify further this point. We start from a description
{\it \`a la} Poland - Sheraga (PS) \cite{Pola66} of the GMO model and use 
a series of exact results on reunion exponents of mutually avoiding walks, 
which show indeed that the denaturation transition is first order. More
surprisingly, the transition turns out to be even sharper than that occurring 
in models where excluded volume effects are fully taken into account. This 
conclusion is corroborated by approximate analytical treatments of excluded 
volume effects due to loop-loop interactions and by a series of numerical 
results obtained for polymer lattice models in $d=2$ and $d=3$.

\begin{figure}[b]
\centerline{\psfig{file=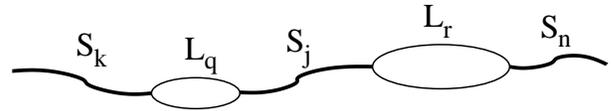,height=1.4cm}}
\vskip 0.2truecm
\caption{In the Poland - Sheraga model the full partition function of the
DNA chain is factorized in terms of products of partition functions $S_k$ 
of bound segments of length $k$ and of partition functions $L_l$ of 
denaturated loops of length $2 l$. Thick lines denote double stranded
segments.
}
\label{FIG01}
\end{figure}  

In the Poland - Sheraga (PS) model \cite{Pola66} the DNA chain is treated 
approximately as a sequence of non-interacting bound double segments and 
denaturated loops. Within this scheme the full partition function $Z_N$ for 
a chain whose constituent strands have lengths $N$, is factorized in terms of 
elementary partition functions for loops and segments (see Fig. \ref{FIG01}). 
In the grand canonical ensemble, where a fugacity $z$ per lattice step is
assigned, the total partition function reads \cite{Fish84}:
\begin{equation}
Z(z) = \sum_N z^N Z_N = \frac{V_0(z)}{1 - L(z) S(z)}
\end{equation}
where $L(z)$ and $S(z)$ are the grand canonical partition functions for loops 
and double segments, respectively. The form of the numerator $V_0(z)$ depends 
on boundary conditions and is not interesting for our purposes. The crucial 
quantity is the loop partition function, which in the most general form 
reads \cite{Fish84}:
\begin{equation}
L(z) = \sum_l \frac{\left( \mu^2 z \right)^l}{(2 l)^c}
\label{loop}
\end{equation}
where $\mu$ is the (non-universal) effective connectivity constant, while
$c$ is a universal exponent. The order of the transition is related to the 
value of $c$ \cite{Fish84}: At the transition temperature (which in the PS 
model corresponds to $\mu^2 z \to 1$) loops are power-law distributed in size 
with probability $P(l) \sim l^{-c}$, thus their average size $\langle l 
\rangle = \sum_l l P(l)$ is finite for $c > 2$, implying a first order 
denaturation, while it diverges for $1 < c \leq 2$ and the transition becomes 
continuous.

For self- and mutually avoiding strands $c$ was estimated analytically in
an extended PS scheme \cite{Kafr00}, where excluded volume effects between a 
self-avoiding loop and the rest of the chain are approximately taken into 
account. This exponent was also obtained directly by Monte Carlo simulations 
\cite{Carl02,Baie02} for polymer lattice models. The agreement between the 
analytical estimates and simulation results for $c$ is very good both in $d=2$ 
and $d=3$ \cite{Kafr00,Carl02,Baie02}.

In the next Sections we will present analytical and numerical estimates of
the exponent $c$ for the GMO model. The former are based on the general
theory of copolymer networks, which is briefly reviewed in the Appendix.

\begin{figure}[b]
\centerline{\psfig{file=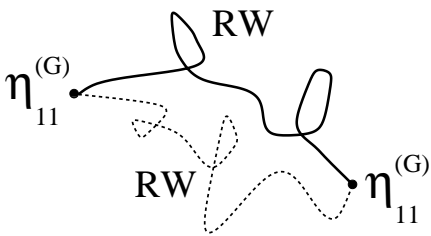,height=3.0cm}}
\vskip 0.2truecm
\caption{
Example of loop in the GMO model. The two strands indicated as dashed and
solid lines are allowed to overlap themselves, but are mutually avoiding.
Equation (\ref{cisolated}) gives the associated reunion exponent.
}
\label{FIG02}
\end{figure}  

\section{Mutually avoiding walks}
\label{sec:maw}

A denaturated loop in the GMO model is 
formed by two mutually avoiding random walks of equal length $l$, with common 
origin and endpoint (see example of Fig. \ref{FIG02}).
The probability that two random walks, with common origin, meet for the 
first time after $l$ steps decays asymptotically as $P(l) \sim l^{-c}$
with \cite{Dupl88,Dupl88b}:
\begin{equation}
c = d \nu - 2 \nu \eta_{\rm 11}^{\rm (G)}.
\label{cisolated}
\end{equation}
Here $d$ is the dimensionality of the embedding space, $\nu = 1/2$ for RWs
and $\eta_{\rm 11}^{\rm (G)}$ denotes the contribution of a vertex from
which the two walks originate (see Fig. \ref{FIG02}).
We follow the notation of Ref. \cite{vonF97} and indicate with
$\eta_{f_1 f_2}^{\rm (G)}$ the exponent associated to a vertex from
which $f_1$ solid and $f_2$ dashed lines depart, at the gaussian fixed
point, which is the relevant one for our purposes. At this fixed point 
solid and dashed lines do not interact with themselves, but they avoid 
the lines of the other group.

Figure \ref{FIG02}, as well as Figs. \ref{FIG03}-\ref{FIG05} below, 
are examples of so called {\it copolymer networks}, 
i.e. networks whose constituents are polymers of different species.
The relevance of these block copolymer networks for a description of
denaturating DNA has been already stressed in the context of models
with full excluded volume effects \cite{Baie02}.
A general network of this type may be made, for instance, by an arbitrary 
mixture of random and self-avoiding walks (SAWs). Any pair of these walks may 
either avoid, or be allowed to cross each other.
As for the homopolymer case, the general entropic exponent for a network with
arbitrary topology follows from those of the constituents vertex exponents
(see Refs.  \cite{Dupl88,Dupl86} for details). The general rule to calculate
the entropic exponent of a network is simple and it is reviewed in the
Appendix. 
Restricting to the case in which all constituent polymers have the
same $\nu$ exponent, one associates to each independent loop a factor
$d \nu$ and to each vertex a factor $-\nu \eta$, with $\eta$ the appropriate
exponent which depends on the number of outgoing legs, and on their type. 
The generalization to an arbitrary mixture of polymer segments with distinct 
$\nu$'s, as for instance RWs and SAWs, is also possible.

Coming back to Eq. (\ref{cisolated}), the vertex exponent for 
one solid and one dashed line is known exactly in $d=2$ \cite{Dupl99}. 
Its value is $\eta_{\rm 11}^{\rm (G)} = -5/4$, which leads to $c=2 + 1/4 >2$.
In higher dimensions one has to resort to $\varepsilon$ expansion
renormalization group results. In $d=3$ resummation techniques yield 
$\eta_{\rm 11}^{\rm (G)} \approx 0.57$ \cite{vonF97}, which implies 
$c \approx 2.07$. 
Thus, both in $d = 2$ and $3$ the reunion exponent is larger than the threshold 
value ($c > 2$), which immediately implies a first order denaturation for the 
GMO model, within the PS picture of non-interacting loops and bound segments.

As a comparison we recall that for an isolated loop made by two self- and
mutually avoiding strands the entropic exponent is \cite{deGe79}:
\begin{equation}
 c = d \nu.
\end{equation}
For a SAW one has $\nu = 3/4$ in $d=2$, which implies $c = 3/2$, while $\nu 
\approx 0.588$ in $d=3$, implying $c \approx 1.76$. Thus, within the PS 
picture, denaturation is a continuous transition ($c < 2$) when full self-
avoidance within a single loop is included \cite{Fish66,Fish84,nota}.

It is also possible to go beyond the PS approximation in the GMO model,
but it is already quite clear at this point that excluded volume effects 
between a single loop and the rest of the chain "localize" even more the 
loops and $c$ can only increase. 
This can be shown explicitly for some simple geometries in which the loop
is embedded. The calculations follow closely those of Ref. \cite{Kafr00}
for the model in which self - avoidance is fully included and are also 
briefly reviewed in the Appendix. We give here only the final results.

\begin{figure}[b]
\centerline{\psfig{file=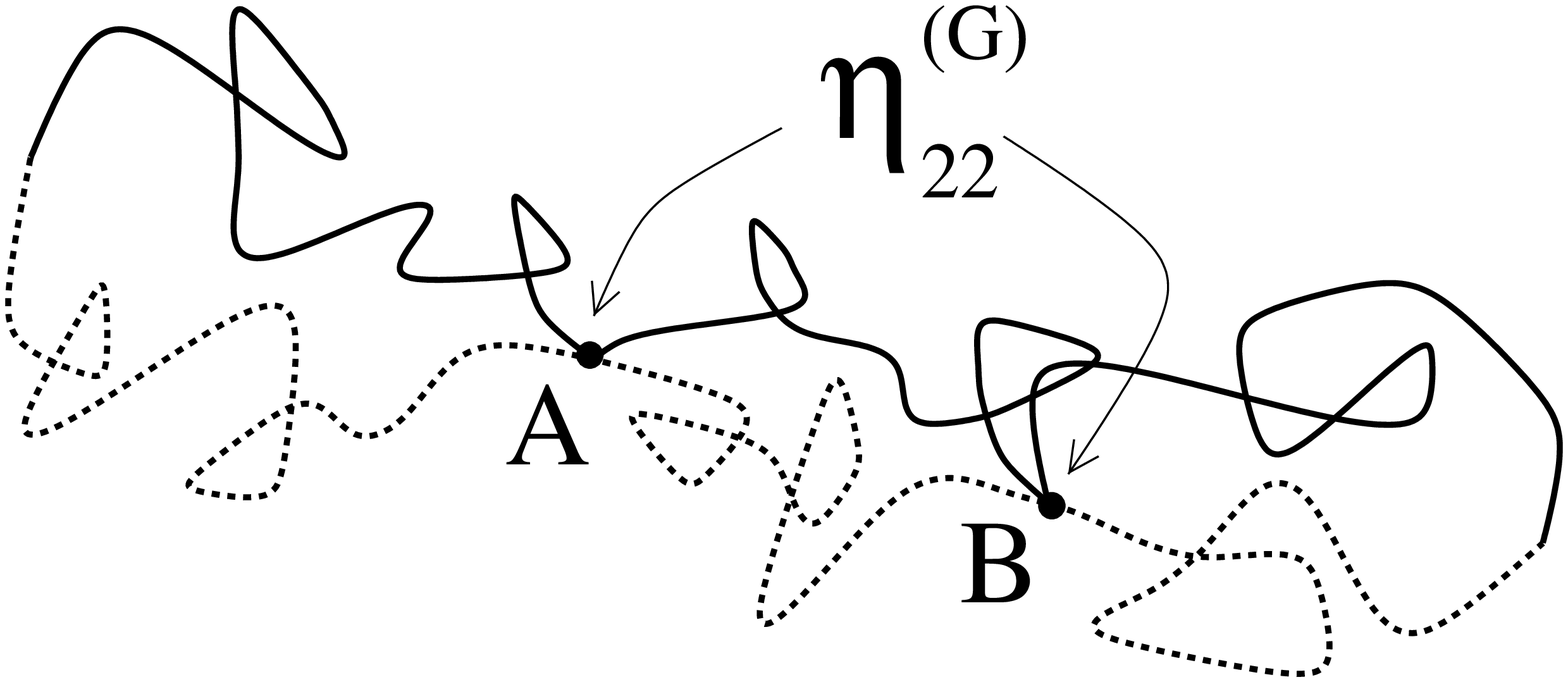,height=3.1cm}}
\vskip 0.2truecm
\caption{
A small loop (between points A and B) embedded between two much larger 
loops is characterized by an exponent given by Eq. (\ref{cinteracting}).
}
\label{FIG03}
\end{figure}

We consider the geometry shown in Fig. \ref{FIG03} in which a loop of 
length $2l$ is connected to two loops of length $N-l$ each. In the limit $l 
\ll N$ the partition functions of the inner loop is still of the type of 
Eq. (\ref{loop}) and factorizes with the partition function of the rest of 
the chain \cite{Kafr00}. As explained in Appendix, the exponent $c$ is however 
modified:
\begin{equation}
c = d \nu - \nu \eta_{\rm 22}^{\rm (G)}.
\label{cinteracting}
\end{equation}

\begin{figure}[b]
\centerline{\psfig{file=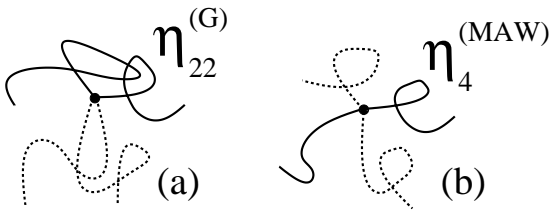,height=2.7cm}}
\vskip 0.2truecm
\caption{In $d=2$ there are two distinct vertices formed by two dashed and 
two solid lines (see text).
}
\label{FIG04}
\end{figure}

In $d=2$ some care has to be taken in considering vertices, as the order of 
walks does matter. This is illustrated in Fig. \ref{FIG04} where the two 
possible distinct types of vertices with four outgoing lines, two solid and 
two dashed, are shown.
For a vertex of type (a) more configurations are possible as neighboring 
lines of the same type are allowed to overlap. In (b), instead, the
distinction between solid and dashed lines becomes irrelevant as 
neighboring lines avoid each other.
In the latter case the lines form a vertex of four mutually avoiding walks 
(MAWs) \cite{vonF02}. A vertex with $L$ outgoing MAWs, has an associated 
entropic exponent \cite{Dupl88,Dupl99}:
\begin{equation}
\eta_L^{\rm (MAW)} = - \frac{4 L^2 - 1}{12}
\end{equation}
Hence, for loops embedded in a geometry as in Fig. \ref{FIG03}, and formed by 
vertices of type (b) one finds $c=d \nu - \nu \eta_4^{\rm (MAW)} 
= 3 + 5/8 = 3.625$.
Let us consider now vertices of type (a) for which we keep the notation 
$\eta_{\rm 22}^{\rm (G)}$ to indicate the associated exponent. Using the
prescriptions of Ref. \cite{Dupl99} one has $\eta_{\rm 22}^{\rm (G)} = 35/12$
and thus from Eq. (\ref{cinteracting}) $c = 2 + 11/24 \approx 2.46$.
Notice that stronger excluded volume effects (as for vertices of type (b))
imply a higher value for $c$, with a quite large difference between the
two estimated values. 
We point out that, obviously, $\eta_{\rm 11}^{\rm (G)} = \eta_2^{\rm (MAW)}$
in all dimensions.

In $d=3$ the above subtleties do not occur.
The vertex exponent $\eta_{\rm 22}^{\rm (G)} \approx -1.81$ was estimated in 
Ref. \cite{vonF97} using $\varepsilon$ expansions and resummation techniques. 
Thus, from Eq. (\ref{cinteracting}), one obtains $c \approx 2.40$.

We consider now another geometry, where a loop is connected to two long double
stranded segments, as illustrated in Fig. \ref{FIG05}. 
Using the same method of Ref. \cite{Kafr00} we find (see Appendix):
\begin{equation}
c = d \nu - 2 \nu \overline{\eta}_{1,2}
\label{c2sigma3}
\end{equation}
where with $\overline{\eta}_{1,2}$ we indicate the exponent associated to a 
vertex where one double, and two single stranded segments join. 
Notice that the double stranded segment is actually a self avoiding walk 
since, if it would overlap itself in some part, the mutual avoidance condition 
between the two constituent strands would be violated.

\begin{figure}[b]
\centerline{\psfig{file=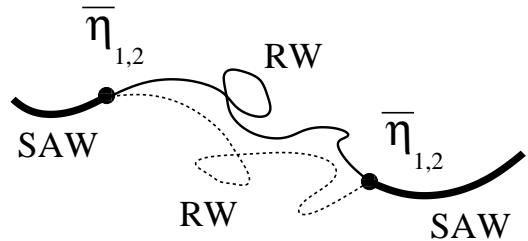,height=3.1cm}}
\vskip 0.2truecm
\caption{
A loop embedded between two long double stranded segments. The latter
follow the self-avoiding walks statistics.
}
\label{FIG05}
\end{figure}

Again in $d=2$ one can use the techniques of Ref. \cite{Dupl99} to find
$\overline{\eta}_{1,2} = -2 - 7/16$ which leads to $c \approx 3.44$. To our 
knowledge, there exists no field theoretical analysis for this type of 
vertices in higher dimensions, thus we are not able to make any predictions 
for $c$ in $d=3$, at present.

Table \ref{TABLE01} summarizes the exponents found in the GMO
model for an isolated loop ($c^{(\rm IL)}$), for a loop interacting with 
other loops ($c^{(\rm LL)}$), as in the geometry of Fig. \ref{FIG03}, and
for a loop interacting with double stranded segments ($c^{(\rm LS)}$), as in
Fig. \ref{FIG05}. For the two dimensional case we reported both values of
$c^{(\rm LL)}$ corresponding to the loops formed by the two types of vertices
of Fig. \ref{FIG04}.
We notice that, as anticipated above, the interaction with other parts of the 
chain has the effect of increasing the loop exponent $c^{(\rm LL)},
c^{(\rm LS)} > c^{(\rm IL)}$ as is the case for all other models studied so 
far \cite{Kafr00,Carl02}. The first order character of the transition in the 
GMO model is therefore strengthened.

It is interesting to compare the exponents obtained in this paper with the 
corresponding ones for the model in which strands are fully self-avoiding 
(FSA). The latter, taken from Ref. \cite{Kafr00}, are given in Table 
\ref{TABLE01}. The comparison between the two models indicates that the
transition in the GMO model is generally sharper (higher $c$). 
This is always true with the exception of the exponent $c^{\rm (LL)}$ in $d=2$
for vertices of the type of Fig. \ref{FIG04}(a) for which we find 
$c^{\rm (LL)} \approx 2.46$, for the GMO model, while $c^{\rm (LL)} \approx
2.69$ in the FSA case. This is due to the fact that in the GMO model for
the type of vertex considered excluded volume effects are not so pronounced 
as walks have considerable freedom to overlap.

\vbox{
\begin{table}[tb]
\caption{ 
Summary of the exponents for the GMO model for an isolated loop (IL) and for 
a loop interacting with neighboring loops (LL) or segments (LS). The results 
for fully self-avoiding (FSA) strands (from Ref. \protect\cite{Kafr00}) are 
also given. The last column shows the numerical estimates of $c$ (those for
the FSA are taken from Ref. \protect\cite{Baie02}).
}
\label{TABLE01}
\vskip 0.2truecm 
\begin{tabular}{c|cccc}
   & $c^{\rm (IL)}$ & $c^{\rm (LL)}$ & $c^{\rm (LS)}$  & c\\
\hline
 GMO (d=2) & 2.25 & 2.46-3.62 & 3.44 & 2.95(5)\\
 FSA (d=2) & 1.50 & 2.69 & 2.41 & 2.46(9) \\
 GMO (d=3) & 2.07 & 2.40 & - & 2.55(5)\\
 FSA (d=3) & 1.76 & 2.22 & 2.11 & 2.18(6) 
\end{tabular}
\end{table}
}

\section{Numerical Results}
\label{sec:num}

We performed a series of 
numerical calculations for a lattice version of the GMO model both in $d=3$ 
and $d=2$ (cubic and square lattices).
We considered two random walks of length $N$, described by the vectors 
$\vec{r}_1 (i)$ and $\vec{r}_2 (i)$ ($i = 0, 1, 2 \ldots N$) which identify 
the positions of the lattice monomers. The walks have common origin 
($\vec{r}_1 (0)= \vec{r}_2 (0)$), and no specific boundary condition is
imposed on the other ends.
While each strand can overlap itself, mutual
avoidance requires that $\vec{r}_1 (i) \neq \vec{r}_2 (j)$ for $i \neq j$.
Overlaps at homologous points are however allowed ($\vec{r}_1 (i) = 
\vec{r}_2 (i)$); these correspond to a bound state of complementary base 
pairs, to which we assign an energy $\varepsilon = -1$. At sufficiently low
temperature ($T$) the walks are fully bound and form an object 
obeying self-avoiding walk statistics.

\begin{figure}[b]
\centerline{\psfig{file=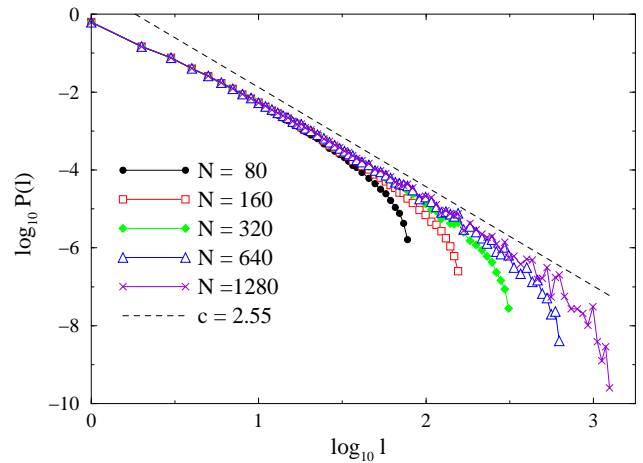,height=6.25cm}}
\vskip 0.2truecm
\caption{Plot of $\log_{\rm 10} P$ vs $\log_{\rm 10} l$ at the estimated 
critical point in $d=3$ and up to $N=1280$. A linear fit of the data yields 
$c = 2.55(5)$.}
\label{FIG06}
\end{figure}

We used the PERM \cite{Gras97} algorithm and computed $P(l)$ the probability 
distribution function (pdf) to find a loop of length $2l$ in the chain. 
While at low temperatures loops are exponentially distributed in size, at the 
critical point the pdf has an algebraic decay $P(l) \sim l^{-c}$, from which 
we can extract the exponent $c$ \cite{Carl02,Baie02}.
Figure \ref{FIG06} shows a log-log plot $P(l)$ vs. $l$ in $d=3$ at the 
estimated critical point ($T_c =0.5181(3)$), for chains of lengths up to 
$N = 1280$. A linear fit of the data yields $c=2.55(5)$, not far from the
$c^{\rm (LL)}$ given by Eq. (\ref{cinteracting}) (see Table \ref{TABLE01}). 
The result confirms 
that the transition in the GMO model is sharper than when self-avoidance is 
fully included. Indeed, in the latter case the most accurate numerical estimate 
of the loop exponent is $c = 2.18(6)$ \cite{Baie02}.

We consider next $P' (r)$, the probability that the distance between homologous
points in the strands $|\vec{r}_1 (k) - \vec{r}_2 (k)|$ equals $r$. This 
quantity was found to decay as a stretched exponential for $r \to \infty$
\cite{Gare01}, from an analysis of the Hamiltonian of Eq. (\ref{hamin}).

\begin{figure}[b]
\centerline{\psfig{file=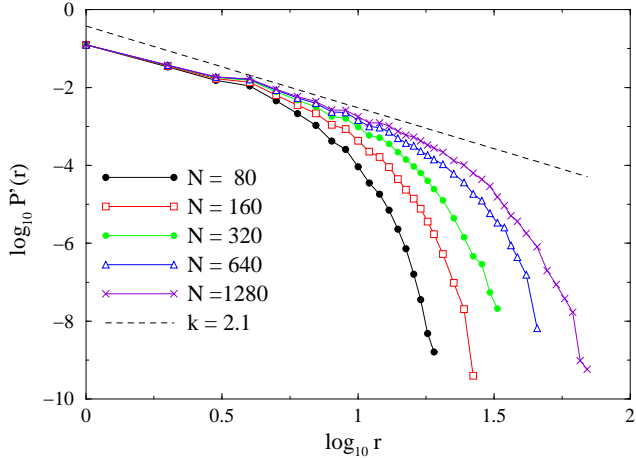,height=6.25cm}}
\vskip 0.2truecm
\caption{
Plot of $\log_{\rm 10} P'$ vs. $\log_{\rm 10} r$. The sets are the same as in 
Fig. \protect\ref{FIG06}. A linear fit yields $k =2.1(1)$.
}
\label{FIG07}
\end{figure}

This conclusion is at odd with the algebraic decay of the loop pdf found here 
for the GMO model, and also in other similar models with a first order 
transition \cite{Carl02}.
For the numerical calculation of $P'(r)$ we registered all distances between 
homologous pairs along the chains $|\vec{r}_1 (k) - \vec{r}_2 (k)|$, starting 
from $k=1$ and up to the highest $k$ for which the two strands are in contact
($\vec{r}_1 (k) = \vec{r}_2 (k)$).
Figure \ref{FIG07} shows a plot of $\log_{\rm 10} P' (r)$ vs. $\log_{\rm 10} r$ 
at the transition temperature in $d = 3$. We cannot fit the data with a 
stretched exponential decay, but by increasing the system size the data 
approach a power law decay $P' (r) \sim r^{-k}$ with $k = 2.1(1)$. 
The fact that $k > 1$ implies that the probability is normalizable. 
This is a confirmation that
the transition is of first order type (see Ref. \cite{deGe79}). 
Notice however, that the algebraic nature of $P'(r)$ implies 
infinite moments $\langle r^n \rangle$ starting from $n=2$. 
We conclude that the exponential decay found in Ref. \cite{Gare01} is an 
artifact of the approximation introduced.

We performed a series of calculations also in the two dimensional case.
At the estimated critical temperature ($T_c = 0.5811(7)$) we find, from the
decay of the probability distributions $P(l)$ and $P'(r)$, the estimates 
$c = 2.95(5)$ and $k = 3.0(1)$, confirming the first order nature of the 
transition. Again denaturation happens to be
sharper than in the case where self-avoidance is fully incorporated, for
which $c \approx 2.46$ \cite{Baie02}. In the PERM algorithm both types 
of vertices of Fig. \ref{FIG04} are generated, thus $c$ cannot be directly 
compared with any of the value obtained analytically in the previous section. 

\section{Conclusion}
\label{sec:concl}

In this paper we studied the model for DNA denaturation recently introduced
by Garel, Monthus and Orland \cite{Gare01}. Our analysis is complementary to 
that of Ref. \cite{Gare01} and it is based on the calculation of the exponent 
$c$ describing the algebraic decay of the probability of denaturated loop 
lengths at the critical point. Using exact results from mutually avoiding walks 
statistics we find that $c > 2$ already for isolated loops, implying that the 
transition is of first order type.
This conclusion is in agreement with the results of Ref. \cite{Gare01}, but 
it is not based on the approximate treatments \cite{Batt02}.

For some specific simple geometries in which a loop is embedded in the chain
we provide analytical estimates of $c$, following ideas from the theory of 
polymer networks \cite{Kafr00,Dupl86}. These calculations show explicitly
that $c$ increases with respect to the isolated loop value, i.e. the transition
robustly maintains its first order character.
More surprising is, at first sight, the fact that the denaturation in the GMO 
model is sharper than that occurring in models where excluded volume 
constraints are fully incorporated (higher $c$), as indicated by our analytical
and numerical estimates of $c$ both in $d=2$ and $3$. This behavior can be 
explained by taking into account that the self-avoidance of the single strands 
somehow contrasts the excluded volume effects between each loop and the rest 
of the molecule. The result is that, when the single strand self-avoidance is 
relaxed, the loop sizes are further reduced, consistently with an increase of 
$c$.

It is somehow remarkable how results of the theory of block copolymer networks,
which are relevant to issues like that of establishing a link between field
theory and multifractal measures (see e.g. Ref. \cite{vonF97}), allow to draw 
reliable predictions on models of the denaturation transition of DNA.

There is another possibility of relaxing the self-avoidance which is somehow 
complementary to that of the GMO model. One can indeed neglect the mutual 
avoidance between the strands, while keeping self-avoidance within each stand.
This possibility was studied in $d=3$ in Ref. \cite{Carl02} and the 
denaturation transition turns out to be continuous ($c < 2$).
Thus, relaxing the self-avoidance constraint in two different ways 
within the single loop, results 
in a softening (as in the latter example) or a sharpening (as in the GMO 
model) of the transition. 

Finally, we also considered the probability distribution of the distances 
between homologous monomers of the two strands. We found that this quantity 
decays as a power law at the transition point, in disagreement with the
stretched exponential behavior found in Ref. \cite{Gare01}, which is probably
an artifact of the approximation introduced there.
The algebraic decay of various quantities at denaturation, even in the case
of first order transitions, is a common feature of other similar models.
Finding the appropriate treatment of the Hamiltonian (\ref{hamin}), which 
provides power-laws decay at the transition point is still an open question.

\section*{acknowledgements}

Useful discussions with David Mukamel are gratefully aknowledged.

\begin{figure}[b]
\centerline{\psfig{file=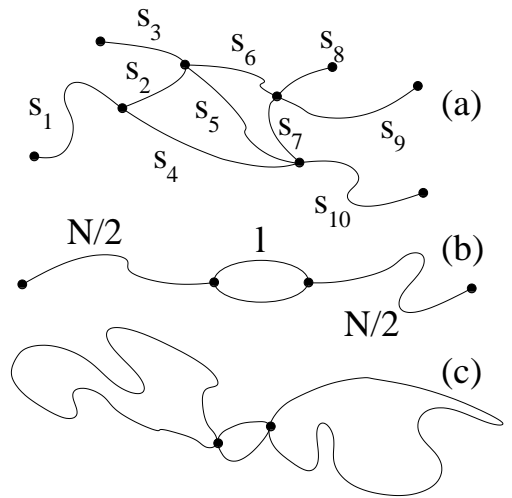,height=6.5cm}}
\vskip 0.2truecm
\caption{Examples of polymer networks with (a) ${\cal L}=2$, $n_1 = 5$,
$n_3 = 2$ and $n_4 = 2$, (b) ${\cal L}=1$, $n_1 = 2$, $n_3 = 2$ and
(c) ${\cal L}=3$, $n_4 = 2$.}
\label{FIG08}
\end{figure}

\section*{Appendix}

We review here some known results for entropic exponents of networks of
arbitrary topology. We focus mainly on networks whose constituent polymers
are all of the same type, and just briefly mention copolymer networks.

Let us consider a network made of self-avoiding walks, all avoiding each 
other, as the one shown in Fig. \ref{FIG08}(a). Let $N = \sum_i s_i$ be the 
total length of the network, where each segment has length $s_i$.
In the asymptotic limit $N \to \infty$, $s_i \to \infty$, the total number 
of configurations for the network scales as \cite{Dupl86}:
\begin{equation}
\Gamma \sim \mu^N N^{\gamma_G - 1} f\left( \frac{s_1}{N}, \frac{s_2}{N} 
\ldots \right)
\label{partfunc}
\end{equation}
where the factor $\mu$ is the effective connectivity constant for the walks
(see e.g. Ref \cite{Vand98}) and $f$ is a scaling function. 
The universal exponent $\gamma_G$ depends only on the number of independent 
loops, ${\cal L}$, and the number of vertices with $k$ outgoing segments $n_k$
as \cite{Dupl86}
\begin{equation}
\gamma_G = 1 - {\cal L} d \nu + \sum_k \nu \eta_k n_k
\label{gammaG}
\end{equation}
where $\eta_k$ are exponents associated to a vertex with $k$ outgoing legs.
Such exponents are known exactly in $d=2$ from conformal invariance 
\cite{Dupl86}, while they have been obtained from renormalization group 
and resummation techniques in $d=3$ \cite{Scha92}.
We notice that it is also customary to use the definition $\sigma_k \equiv 
\nu \eta_k$.

This theory was recently applied to the denaturation problem \cite{Kafr00}.
In this case one considers network geometries as that of Fig. \ref{FIG08}(b),
in which a loop of total length $2l$ is connected to two polymers of lengths 
$N/2$ each. Following the above notation one has ${\cal L}=1$, $n_1=2$ and 
$n_3=2$ and thus the total number of configurations scale as:
\begin{equation}
\Gamma \sim \mu^N N^{\tilde \gamma - 1} f\left(\frac{l}{N} \right)
\label{eq11}
\end{equation}
with ${\tilde \gamma} = 1 -d \nu + 2 \nu ( \eta_1 + \eta_3 )$. 
Now in the limit $l \ll N$ one should recover the partition function for a
self-avoiding walk of total length $N$ which has an entropic exponent 
$\gamma = 1 + 2 \nu \eta_1$. This requires that:
\begin{equation}
f(x) \sim x^{2 \nu \eta_3 - d \nu}
\end{equation}
in the limit $x \to 0$. Thus, loops of length $l$ embedded within a long chain 
of length $N$ ($ \gg l$) are distributed according to the probability $P(l) 
\sim l^{-c}$, with \cite{Kafr00}:
\begin{equation}
c =  d \nu - 2 \nu \eta_3.
\label{sigma3}
\end{equation}
 
As second example we consider a loop of length $2l$ embedded between two long 
loops of total length $2N$, as depicted in Fig. \ref{FIG08}(b). The total 
number of configurations is still given by Eq. (\ref{eq11}) with ${\tilde 
\gamma} = 1 - 3 d \nu + 2 \nu \eta_4$, as the network has now ${\cal L} = 3$ 
independent loops and two vertices with four outgoing legs, i.e. $n_4 = 2$. 
In the limit $l \ll N$ one should find the partition function of two loops 
joined at a single vertex.
Matching the scaling function in this limit yields
\begin{equation}
f(x) \sim x^{\nu \eta_4 - d \nu}
\end{equation}
for $x \to 0$. One finds thus that loops are distributed according to the 
probability $P(l) \sim l^{-c}$, with \cite{Kafr00}:
\begin{equation}
c = d \nu - \nu \eta_4.
\label{sigma4}
\end{equation}

Finally, this theory can be generalized to the case of networks with more 
than one type of polymer, as for {\it copolymer} networks. Eq. (\ref{gammaG})
can be generalized easily in the case that the constituting polymers have all
the same metric exponents (for instance in the case of mutually avoiding 
random walks). The only difference is that now vertex exponents will not 
depend only on the number of outgoing legs, but also on the type of polymers
forming the vertex and, in two dimensions, also on the order.
For instance, for a network made of two types of polymers, which are random
walks and mutually avoiding with a geometry of Fig. \ref{FIG03} one arrives 
to a $c$ exponent in the limit of an inner loop much shorter than the rest of
the chain as given by Eq. (\ref{cinteracting}), which is the copolymer network 
counterpart of Eq. (\ref{sigma4}). 
Notice that while for homopolymers $\eta_{\rm 2} = 0$, in general a vertex with
two outgoing polymers of different type will have a non-vanishing exponent (as 
is the case of the $\eta_{\rm 11}^{\rm (G)}$ introduced in Fig. \ref{FIG02}).

Finally if polymers constituting the inhomogeneous network will have different
$\nu$ exponent, as for networks formed by a mixture of random and self-avoiding
walks than Eq. (\ref{partfunc}) is still valid, but in a form involving the
radii of gyrations $R_i$ of the constituent polymers, where 
$R_i \sim s_i^{\nu}$, for a segment of length $s_i$.

\end{multicols}


\begin{references}
\bibitem{nota} Notice that denaturation becomes first order for $d > 4$,
since $\nu =1/2$ and thus $c = d \nu = d/2 > 2$.
\end{references}
\end{document}